\documentclass[preprint,showpacs]{revtex4}
\usepackage{graphicx}
\usepackage{color}
\usepackage{amssymb,amsmath}
\usepackage{hyperref}

\begin{document}

\title{A physical limitation of the Wigner ``distribution'' function in transport}

\author{Ioan B\^aldea}
\email{ioan.baldea@pci.uni-heidelberg.de}
\altaffiliation[Also at ]{National Institute for Lasers, Plasma, 
and Radiation Physics, ISS, POB MG-23, RO 077125 Bucharest, Romania.}
\author{Horst K\"oppel} 
\affiliation{Theoretische Chemie,
Physikalisch-Chemisches Institut, Universit\"{a}t Heidelberg, Im
Neuenheimer Feld 229, D-69120 Heidelberg, Germany}
\begin{abstract}
We present an example revealing that the sign of the ``momentum'' $P$ of 
the Wigner ``distribution'' function $f(q, P)$ is not necessarily associated with the direction 
of motion in the real world. This aspect, which 
is not related to the well known limitation of 
the Wigner function that traces back to the Heisenberg's uncertainty principle,
is particularly relevant in transport studies, wherein it is helpful to distinguish between 
electrons flowing from electrodes into devices and vice versa.
\end{abstract}

\date{\today}
\maketitle

Recently, we critically analyzed \cite{Baldea:2008b,Baldea:2009c,Baldea:2010g,Baldea:2011a}
the manner in which the Wigner ``distribution'' function was used in studies on
molecular transport relying upon finite isolated clusters \cite{DelaneyGreer:04a}.
These studies made us aware of a limitation of using the Wigner 
function as a true momentum distribution function 
for transport, which we could not find in the literature and want to present here.

The Wigner function $f(q, P)$ is employed in many physical studies, including transport's,
in spite of its physical limitations. 
The limitation known from textbooks \cite{mahan,Datta:97} 
traces back to the Heisenberg's uncertainty principle.
The Wigner ``distribution'' 
function can be negative and should be not interpreted as a probability distribution,
but rather as ``one step in the calculation {\ldots} never the last step, since''  
is not measurable but ``is used to calculate other quantities that 
can be measured {\ldots} the 
particle density and current'' and ``no problems are encountered as long as one avoids 
interpreting $f$ as a probability density'' (quotations from 
ch.~3.7, p.~203 of Ref.~\onlinecite{mahan}). 

In transport, it is helpful
to distinguish between incoming and outgoing electrons, i.~e., flowing from 
electrodes into devices and from devices into electrodes \cite{Frensley:90}. 
To this aim,
it is necessary to use a physical property whose sign enables to
indubitably assess that electrons are, say, left- or right-moving.
The averages of the particle (probability) current operator
\begin{equation}
\hat{\j}(x) = -\frac{i}{2 m\hbar}
\left[ 
\hat{\psi}^{\dagger}(x) \frac{\partial \hat{\psi}(x)}{\partial x} - 
\frac{\partial \hat{\psi}^{\dagger}(x)}{\partial x} \hat{\psi}(x) 
\right] ,
\end{equation}
or of the 
{\em physical} momentum $\hat{P}_x = -(i/\hbar)\,(\partial / \partial x )$ 
do represent such properties. 
Above, 
$\hat{\psi}(x)$ and $\hat{\psi}^{\dagger}(x)$ 
are annihilation and creation field operators for
(spinless) electrons moving in one dimension, 
and $m$ stands for electron's mass.
To see whether the momentum ``variable'' $P$ of the Wigner function 
\begin{equation}
\label{eq-wf}
f(q, P) \equiv \frac{1}{N}\int d\,r  \ e^{-i P r} 
\left\langle \Psi\left\vert \hat{\psi}^{\dagger}(x - r/2) \hat{\psi}(x + r/2) 
\right\vert\Psi\right\rangle,
\end{equation}
justifies to speak of left- or right-moving
electrons depending on the sign of $P$,
let us consider the ground state $\vert \Psi\rangle$ of $N$
noninteracting electrons confined within a one-dimensional
square well of width $L$ and infinite height.
Electrons occupy energy levels $\hbar^2 \kappa^2/(2 m)$, whose
single-electron wave functions
\begin{equation}
\varphi_{\kappa}(x) = (2/L)^{1/2} \sin(\kappa x) = 
i (2 L)^{-1/2}\left( e^{-i\kappa x } - e^{i \kappa x} \right) ,
\end{equation}
with $\kappa \to \kappa_n = \pi n/L$, $n=1,2,3,{\ldots}$.
In the ground state $\vert \Psi\rangle$, 
the lowest $N$ levels are occupied up to the Fermi ``momentum''
$p_F = \hbar k_F = \pi \hbar N/L$.
Computing the Wigner function of this system is straightforward
\begin{equation}
f(q, P) = \sum_{\kappa_n \leq \kappa_F}
\int_{x \pm r/2 \leq 0} d\,r e^{-i P r}
\varphi_{\kappa_n}^{\ast}\left(x-\frac{r}{2}\right)\varphi_{\kappa_n}\left(x+\frac{r}{2}\right);
\end{equation}
see, e.~g., Ref.~\onlinecite{mahan}, ch.~3.7, pp.~202-203.
\begin{figure}[h]
\centerline{\includegraphics[width=0.45\textwidth,angle=0]{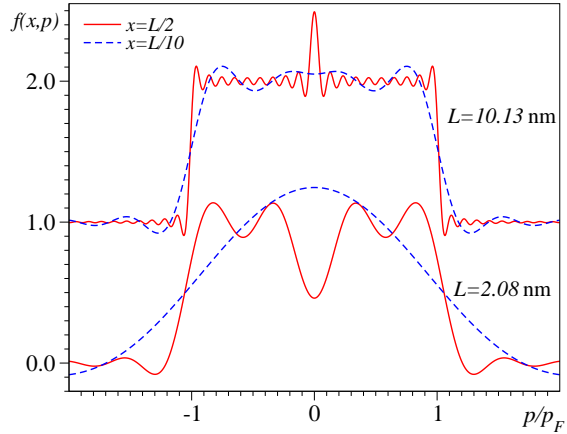}}
\caption{\label{fig:wf} (Color online)
Wigner function for two sizes $L$ computed at two points $x$ indicated in the legend.
Gold's Fermi wave vector $p_F/\hbar=12$\,nm$^{-1}$ is used.}
\end{figure}

One might think that one could use the Wigner function as if it were
a distribution function in cases where its shape resembles a Fermi distribution.
Let us inspect the curves for $f(q, P)$ computed as indicated above and
presented in Fig.~\ref{fig:wf}.
In fact, at smaller sizes
(close to the linear size of the Au$_{13}$-clusters used in molecular 
Wigner transport studies \cite{DelaneyGreer:04a})
the Wigner function does not bear much resemblance to a Fermi function (the lower curves of Fig.~\ref{fig:wf}),
At larger sizes (much larger than those one could hope to tackle within ab initio calculations
to correlated molecules, for which such a Wigner-transport approach 
\cite{DelaneyGreer:04a} was conceived)
the curves (the upper part of Fig.~\ref{fig:wf}) become more similar to a step function,
and one may think
that this is encouraging. In reality, the contrary is true:
as visible in Fig.~\ref{fig:wf}, \emph{mathematically} one can calculate the Wigner function
for positive and negative ``momentum'' variables $P$ \emph{separately}.
However, this mathematical separation does not reflect a physical reality:
for any single-particle eigenstate $\kappa$
the electron momentum vanishes
\begin{equation}
P_{\kappa} \equiv -i \hbar
\int d\,x \varphi_{\kappa}^{\ast}(x) \left( \partial/\partial x\right)
\varphi_{\kappa}(x) = 0 ;
\end{equation}
left- and right-traveling waves are entangled with equal weight, and
one cannot speak of single-particle
eigenstates representing left- or right-moving electrons only because
Wigner functions with positive or negative $P$-arguments can be computed.

This represents a further limitation of the usefulness of the Wigner function,
not related to the Heisenberg's principle,
which is particularly relevant for transport. The current has a direction,
and if one wants to unambiguously specify this direction, the Wigner function $f(q, P)$
is inappropriate;
a Wigner function with negative (positive) ``momentum'', $ P < 0$ ($ P > 0$),
does not imply that left- and right-moving particles exist in the real physical world.

So, using $f(q_L, P>0)$ and $f(q_R, P<0)$ as if they were true 
momentum distributions of 
incoming electrons, as done in Wigner approaches of molecular transport 
based on finite isolated clusters \cite{DelaneyGreer:04a} is not justified
in quantum mechanics. The above example demonstrates that, indeed, 
the textbook's warning mentioned in the beginning of this 
section is pertinent.
\section*{Acknowledgment}
The financial support provided by the Deu\-tsche For\-schungs\-ge\-mein\-schaft 
is gratefully acknowledged.
\end{document}